\pdfoutput=1

\documentclass[11pt]{article}

\usepackage[final]{acl}

\usepackage{times}
\usepackage{latexsym}
\usepackage{tipa}
\usepackage[table,xcdraw]{xcolor}
\usepackage{tabularx}

\usepackage[T1]{fontenc}

\usepackage[utf8]{inputenc}

\usepackage{microtype}

\usepackage{inconsolata}

\usepackage{graphicx}

\usepackage{diagbox}
\title{Read to Hear: A Zero-Shot Pronunciation Assessment Using Textual Descriptions and LLMs}

\author{Yu-Wen Chen \quad Melody Ma \quad Julia Hirschberg \\
Department of Computer Science, Columbia University, United States \\
\texttt{yuwchen@cs.columbia.edu}}

\begin{document}
\maketitle
\begin{abstract}
Automatic pronunciation assessment is typically performed by acoustic models trained on audio-score pairs. Although effective, these systems provide only numerical scores, without the information needed to help learners understand their errors. Meanwhile, large language models (LLMs) have proven effective in supporting language learning, but their potential for assessing pronunciation remains unexplored. In this work, we introduce TextPA, a zero-shot, Textual description-based Pronunciation Assessment approach. TextPA utilizes human-readable representations of speech signals, which are fed into an LLM to assess pronunciation accuracy and fluency, while also providing reasoning behind the assigned scores. Finally, a phoneme sequence match scoring method is used to refine the accuracy scores. Our work highlights a previously overlooked direction for pronunciation assessment. Instead of relying on supervised training with audio-score examples, we exploit the rich pronunciation knowledge embedded in written text. Experimental results show that our approach is both cost-efficient and competitive in performance. Furthermore, TextPA significantly improves the performance of conventional audio-score-trained models on out-of-domain data by offering a complementary perspective.

\end{abstract}

\section{Introduction}

Automatic pronunciation assessment offers an alternative to traditional language instruction by providing learners with accessible, scalable, and timely feedback on their speaking abilities. Most prior work in this area relies on supervised learning: collecting speech recordings annotated with pronunciation scores from human instructors and training acoustic models to assess proficiency scores~\cite{chen2023multipa, gong2022transformer}. Although effective, models trained on audio-score pairs provide only numerical scores, offering little insight into why a particular score was assigned. Collecting more informative and descriptive feedback, such as detailed comments from human raters, can be time-consuming and expensive.

Recently, Large Language Models (LLMs) have gained popularity for their ability to generate natural, context-aware responses. We propose that this generative capability can be leveraged to produce explainable feedback in pronunciation assessment, going beyond simple scoring. Furthermore, LLMs have demonstrated the potential to provide valuable insights into language learning~\cite{c2023impact}. Most studies focus on the use of LLMs in writing tasks~\cite{lo2024exploring}, but LLMs also capture knowledge of language speaking, as humans have documented their knowledge about pronunciation in written form to facilitate sharing and teaching. In addition, previous studies have shown that LLMs, such as GPT, have the potential to interpret textual descriptions of speech signals. In~\cite{wang2023assessing}, researchers wrote the pause durations in a sentence -- e.g., “it (\textless 10 ms) is (\textless 10 ms) nothing (10 ms–50 ms) like (\textless 10 ms) this,” -- and put the sentence into GPT to assess whether the pauses are correct. However, this study focused only on detecting inappropriate pauses using duration information, without exploring the ability of LLMs to interpret other key dimensions of pronunciation, such as articulation or intonation.

To bridge the gap between the textual understanding of LLMs and the physical acoustic signal, audio-language models (ALMs)~\cite{elizalde2023clap, tang2023salmonn, chu2023qwen} have emerged. ALMs integrate audio and text by encoding audio into audio tokens, which are then processed by the LLM with text tokens. However, most open-source ALMs are pre-trained on audio captioning or speech recognition datasets and show limited ability to assess speech without fine-tuning~\cite{deshmukh2024pam, wang2025enabling}. In addition, due to computational constraints, these studies used smaller LLMs (e.g., 7B or 13B Llama), limiting their ability to fully leverage LLM capabilities. On the other hand, commercial large ALMs such as GPT-audio and Gemini-audio have demonstrated the potential for pronunciation assessment in zero-shot settings~\cite{wang2025exploring}, but these ALMs are costly to operate with an audio input. Since audio tokens are much more expensive than text tokens\footnote{For example, the OpenAI \textit{GPT-4o-mini-audio} model charges \$10.00 per 1M audio tokens, compared to \$0.15 per 1M text tokens (as of April 2025).} and the number of audio tokens generated from a speech signal can be much greater than the number of text tokens in its corresponding transcript, using a large ALM with audio inputs is considerably more expensive than using LLM with text inputs.

Therefore, we explore an alternative method to bridge the gap between LLM's textual knowledge and physical speech signals. Instead of relying on audio tokens, our method uses the existing capabilities of LLMs by selecting text-based acoustic descriptors common in written text. Pre-trained acoustic models are used to generate these, including transcripts, phoneme sequences (in both International Phonetic Alphabet (IPA) and CMU Pronouncing Dictionary (CMU) formats), and pause durations. The descriptors are provided as input to LLMs for pronunciation assessment. Lastly, we incorporate a similarity score between the recognized IPA sequence and the canonical IPA sequence mapped from the transcript to improve the assessment of pronunciation accuracy.

The contributions of this work are summarized as follows: (1) We propose TextPA, a zero-shot pronunciation assessment model that uses textual descriptions of speech signals. (2) Our method produces interpretable and explainable feedback, unlike conventional pronunciation assessment systems that yield only numeric scores. In addition, incorporating TextPA enhances the performance of an audio-score-trained model on out-of-domain data. (3) Compared to large ALMs, our approach significantly reduces API costs while delivering competitive or superior assessment performance.

\section{TextPA}
\begin{figure*}[]
\centering
  \includegraphics[scale=0.9]{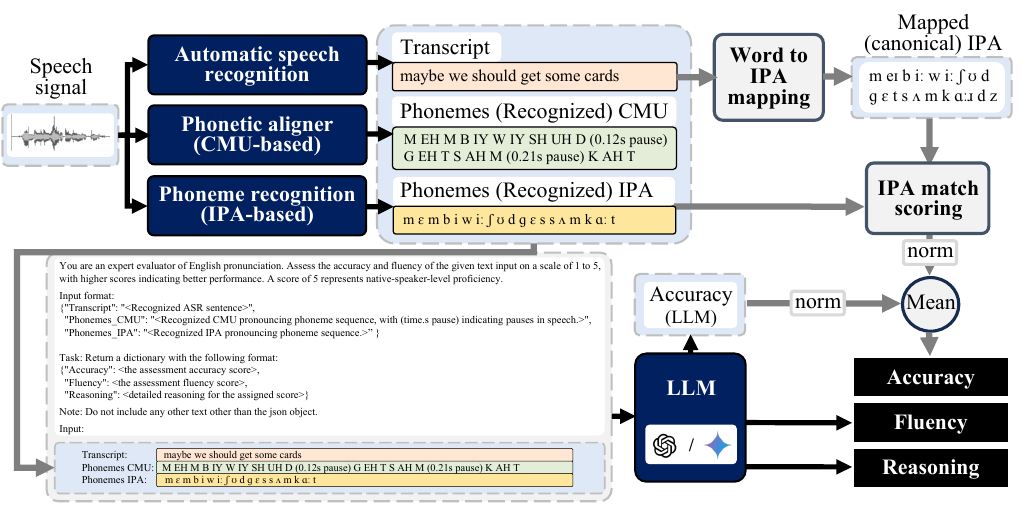} \hfill
  \caption {An overview of TextPA.}
    \label{fig:TextPA}
\end{figure*}

To assess English pronunciation in terms of accuracy and fluency, textual acoustic cues are extracted using a set of pre-trained models: the transcript is obtained from an automatic speech recognition (ASR) model; pause information and the recognized CMU sequence are derived from a phonetic aligner; and the IPA phoneme sequence is generated using a phoneme recognition model. These textual representations are then provided as input to an LLM, which is prompted to assess the pronunciation and produce both accuracy and fluency scores, along with the reasoning behind its evaluations. Lastly, IPA match scoring is introduced to further refine the accuracy score. Figure~\ref{fig:TextPA} presents an overview of TextPA, which operates in a zero-shot setting by leveraging pre-trained acoustic models and LLMs, and thus does not require audio-score paired pronunciation data for training.

\subsection{Textual Acoustic Cues for LLM Input}

\subsubsection{Transcript}

\begin{figure}[t]
\centering
\includegraphics[scale=0.79]{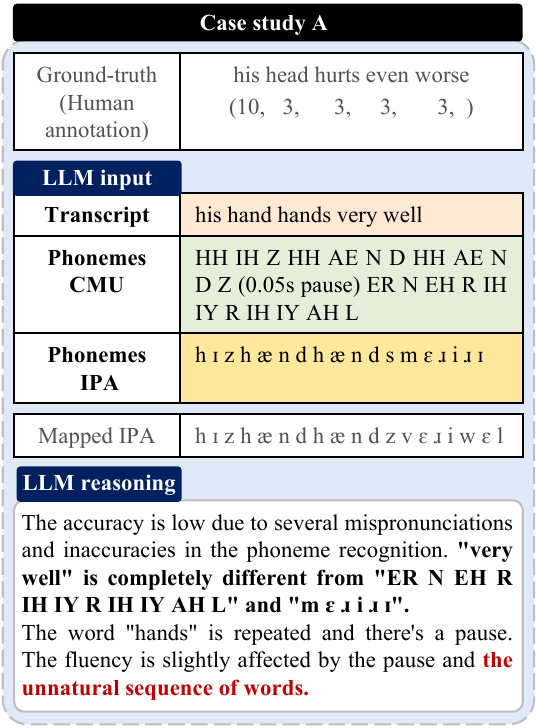}
  \caption{Case study A. Due to the inaccurate pronunciation, the ASR model produced unnatural word sequences, which in turn signaled to the LLM that the pronunciation was flawed. }
  \label{fig:case_A}
\end{figure}

A transcript lacking semantic coherence may result from inaccurate recognition due to poor pronunciation. Repeated words within a sequence or filler words such as \textit{“hmm,”} can indicate a lack of fluency. In Case study A (Figure~\ref{fig:case_A}), the speaker is told to say \textit{“his head hurts even worse,”} but their pronunciation is highly inaccurate. Except for \textit{"His."}, all other words received only \textit{3} out of \textit{10} points. Due to poor pronunciation and lack of fluency, the ASR model produced an inaccurate transcript (i.e., \textit{“His hand hands very well”}) which is semantically incoherent, signaling low pronunciation proficiency for the LLM, as reflected in its reasoning. However, since ASR model is designed to recognize words rather than analyze pronunciation, it may automatically correct inaccurately pronounced words to produce a semantically coherent sentence. For example, in Figure~\ref{fig:case_B}, the speaker is instructed to say \textit{“maybe we should get some cake”} but mispronounced \textit{“cake.”} Although the pronunciation is inaccurate, the ASR transcript (\textit{“maybe we should get some cards,”}) is still semantically reasonable. As a result, the transcript alone is insufficient to reveal the finer details of articulation. To address this, we incorporate the IPA and CMU phoneme sequences that explicitly represent spoken sounds.

\subsubsection{Recognized IPA and CMU Phoneme Sequence}
IPA, widely used in linguistics, dictionaries, and language education materials, is a standardized phonetic notation system that represents the sounds of spoken language using a consistent set of symbols. Each symbol corresponds to a specific speech sound, providing a one-to-one mapping between sound and notation. The CMU phoneme sequence is a phonetic transcription format based on the Carnegie Mellon University Pronouncing Dictionary (CMUdict). Unlike IPA, which is universal in language and more fine-grained, CMU uses a simplified set of phonemes tailored for American English, which is widely used in speech processing applications due to its compatibility with ASR systems and phoneme-based models. Because both representations are widely used, LLMs trained on extensive text corpora have encountered and internalized the mapping between IPA and CMU phoneme annotations and the word. For example, in Case study B (Figure~\ref{fig:case_B}), by comparing the recognized IPA and CMU sequences, the LLM identifies that the word “cards” may have been mispronounced and uses this information to assess pronunciation accuracy. It can align transcript words with the corresponding phoneme sequences even when word boundaries are not explicitly marked. We also embed pause information from the phonetic aligner into the recognized CMU phoneme sequence. Pauses are annotated in an easily interpretable format, e.g. \textit{“D (0.12s pause) G”} indicates a \textit{0.12-second pause} between the phones \textit{“D”} and \textit{“G”}. As shown in Case study B (Figure~\ref{fig:case_B}), the LLM leverages this pause information when reasoning about the speaker’s fluency.

\subsection{IPA Match Scoring}
\begin{figure}[t]
\centering
\includegraphics[scale=0.79]{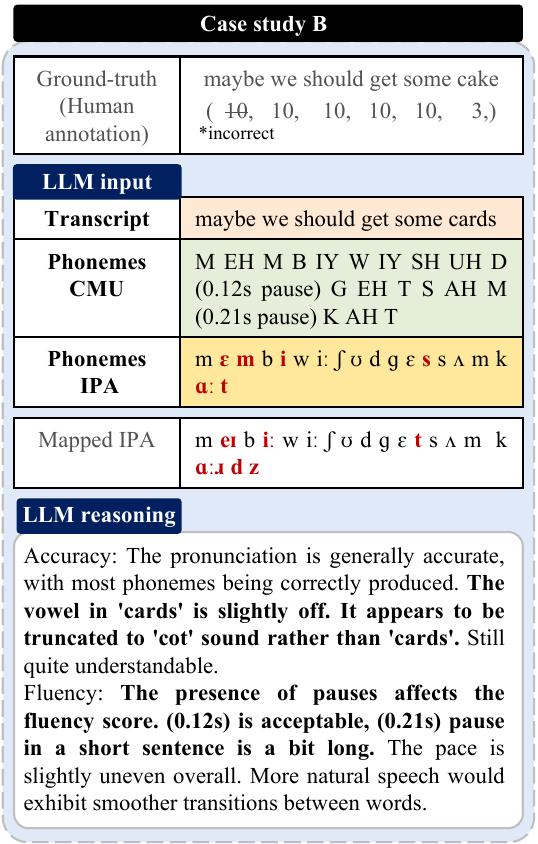}
  \caption{Case study B. The transcript is semantically reasonable despite inaccurate pronunciation. To address this, the mismatch between the recognized phoneme sequence and the transcript provides the LLM with insight into potential articulation inaccuracies. The mapped IPA (i.e., the canonical IPA of the transcript) is shown for reference and is not provided as input to the LLM.}
  \label{fig:case_B}
\end{figure}

To assess pronunciation, the LLM internally maps each word in the transcript to its canonical phoneme sequence and compares it with the provided recognized phoneme sequence. Although LLMs are capable of this, as shown in Case study B (Figure~\ref{fig:case_B}) where the model correctly identifies the mispronunciation of the word \textit{“cards”}, they may still overlook some errors. For example, in the same case, a discrepancy is observed between the canonical phoneme sequence for the word \textit{“maybe”} (\textipa{m eI b i:} / M EY B IY) and the recognized sequence (\textipa{m E m b i} / M EH M B IY), indicating inaccurate pronunciation. Although the human annotation assigns a score of 10 out of 10 to the pronunciation accuracy of \textit{“maybe”}, our manual inspection suggests that the word is not clearly articulated. However, the LLM does not reflect this error in its reasoning.

To further refine accuracy assessment, we introduce IPA match scoring, which measures the similarity between the recognized and canonical IPA sequences and uses this as an indicator of pronunciation accuracy. We use IPA instead of CMU because IPA offers more fine-grained phonetic detail. In addition, our empirical results suggest that match scoring using IPA consistently outperforms scoring with CMU phonemes. To perform IPA match scoring, the transcript is first mapped to its canonical IPA phoneme sequence using a pronunciation dictionary. We then compute the similarity between the transcript-mapped and the recognized IPA sequences using the Smith-Waterman algorithm~\cite{smith1981identification}. This algorithm is a dynamic programming method originally developed for local alignment of biological sequences such as DNA or proteins. It identifies the most similar subsequences between two inputs, making it well-suited for phoneme sequence comparison, which consists of subsequences representing different words. Since the IPA-matched scores and the LLM-generated accuracy scores operate on different scales, we apply min-max normalization across the test set to align them within a common range. The final accuracy score is then calculated as the average of the normalized Smith-Waterman similarity and the normalized LLM-generated score.

\section{Experimental Setup}

\subsection{Data and Evaluation Metric}
We evaluated TextPA on the open-source Speechocean762~\cite{zhang2021speechocean762} and MultiPA~\cite{chen2023multipa} datasets~\footnote{License: Attribution 4.0 International (CC BY 4.0)}, both of which focus on English speech produced by native Mandarin speakers. The Speechocean762 (abbreviated as Speechocean) dataset consists of 5,000 utterances spoken by 250 speakers, with annotations at the sentence, word, and phoneme levels. In this study, we focus on sentence-level accuracy, fluency, and prosody. The utterances in Speechocean are scripted. Participants were instructed to read predefined sentences, making the ground-truth transcript available. However, our method operates without the need for ground-truth information. Most sentences in Speechocean are short, as shown in Figure~\ref{tab:dataset_example},~\ref{fig:case_A}, and ~\ref{fig:case_B}, with corresponding audio durations ranging from 2 to 20 seconds. Since TextPA requires no training, we used only the Speechocean test set, which contains 2,500 utterances.

The MultiPA data contains 50 audio clips, each ranging from 10 to 20 seconds in duration, collected from \textasciitilde20 anonymous users interacting with a dialogue-based chatbot. Unlike Speechocean, where speakers are asked to read predefined sentences, MultiPA data captures open-ended responses, allowing learners to speak freely or answer questions. This allows for a more authentic assessment of learners’ speaking abilities. Table~\ref{tab:dataset_example} shows example transcriptions from both datasets.
\begin{table}[htbp!]
\centering
  \includegraphics[scale=0.77]{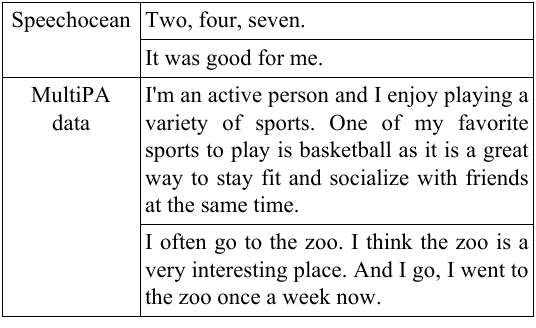} 
  \caption{Example transcriptions from Speechocean and MultiPA. Speechocean consists of relatively short, scripted utterances from read-aloud tasks, whereas MultiPA data captures open-ended, conversational speech.}
  \label{tab:dataset_example}
\end{table}
We use the Pearson correlation coefficient (PCC) as the main evaluation metric since it has often been used in prior studies and provides better interpretability when comparing performance on different datasets. 

\subsection{Implementation Details}

We use Whisper~\cite{radford2023robust} (\textit{large-v3-en}) for transcription, the model from~\cite{xu2021simple}\footnote{\url{https://huggingface.co/facebook/wav2vec2-lv-60-espeak-cv-ft}} for IPA sequence, Charsiu~\cite{zhu2022phone} predictive aligner for CMU sequence, and Phonemize~\cite{Bernard2021}\footnote{\textit{EspeakBackend("en-us")}} for word-to-IPA mapping. Acoustic models were run on an NVIDIA RTX 4500 GPU. The LLMs use default API settings, and results are from a single run.

\section{Results}

\subsection{Performance on Free-speech}
\label{sec:performance_on_free_speech}

Table~\ref{tab:performance_multiPA} shows the performance on MultiPA data. We compare TextPA with different LLM back-ends. Since TextPA \textit{(gpt-4o-mini)} outperforms TextPA \textit{(gemini-2.0-flash)}, we used GPT-4o-mini-audio for the performance comparison. Results suggest that the proposed TextPA outperforms GPT-4o-mini-audio in assessing pronunciation, achieving better performance in both accuracy and fluency. We also compare performance with the MultiPA model~\cite{chen2023multipa}, an acoustic model trained on Speechocean. Results show that the proposed TextPA achieves higher accuracy and provides competitive fluency assessment, showing the effectiveness of TextPA in a zero-shot setting. 

We evaluate the performance of combining the MultiPA and TextPA models. To account for differences in the scale of their prediction scores, we first apply min-max normalization to each model's outputs. The final prediction is obtained by averaging the normalized scores. Despite the simplicity of this fusion strategy, the combined model achieves notable performance improvement over using either model alone. This improvement is likely due to the distinct sources of information. MultiPA is trained on paired audio-score data, learning directly from acoustic examples, whereas TextPA operates solely on text and leverages prior knowledge about pronunciation assessment. Differing approaches offer diverse perspectives, enabling the combined system to achieve improved performance.

Due to the limited amount of paired audio-score pronunciation data, MultiPA may have difficulty accurately assessing words that were not encountered during training. In contrast, TextPA has access to a much broader vocabulary, leading to higher performance on accuracy assessment. However, because MultiPA analyzes raw audio recordings, it can capture acoustic cues such as detailed phone-level durations or pitch variations. These cues are typically not represented in written descriptions or are difficult to capture accurately in text, making them challenging for LLMs to interpret. In fact, we also explore the LLM's ability to assess prosody using ToBI annotations~\cite{beckman1994tobi} which offer a text-based representation of tonal patterns and phrase boundaries. However, the LLM appears to struggle with assessing prosody by accurately interpreting these annotations, even when given explicit instructions (see the Appendix~\ref{sec:appendix} for details). In essence, the two approaches provide complementary advantages on the assessment task, and combining them could be beneficial by leveraging the strengths of both.


\begin{table}[htbp!]
\centering
\renewcommand{\arraystretch}{1.1}
\begin{tabular}{|c|c|c|}
\hline & Accuracy       & Fluency        \\ \hline
\begin{tabular}[c]{@{}c@{}}TextPA\\ \textit{(gemini-2.0-flash)}\end{tabular} & 0.697          & 0.557          \\ \hline
\begin{tabular}[c]{@{}c@{}}TextPA\\ \textit{(gpt-4o-mini)}\end{tabular}      & \textbf{0.728} & 0.650          \\ \hline \hline
GPT-4o-mini-audio                                                   & 0.674          & 0.648          \\ \hline
MultiPA model                                                       & 0.618          & \textbf{0.683} \\ \hline \hline
\begin{tabular}[c]{@{}c@{}}MultiPA model +\\ TextPA \textit{(gpt-4o-mini})\end{tabular} & \textbf{0.769} & \textbf{0.784} \\ \hline
\end{tabular}
  \caption{Model performance on MultiPA data. Note that MultiPA model was trained on Speechocean.}
  \label{tab:performance_multiPA}
\end{table}

\subsection{Performance on Scripted Utterances}

Table~\ref{tab:performance_speechocean} shows the performance on Speechocean. We first compare the performance of TextPA using different LLM back-ends. Results indicate that \textit{gemini-2.0-flash} outperforms \textit{gpt-4o-mini}; therefore, we conducted another experiment using Gemini-2.0-flash-audio for our performance comparison. In contrast to its strong performance on the MultiPA dataset, TextPA performs relatively poorly on Speechocean. This discrepancy might arise from fundamental differences between the datasets. Speechocean consists of shorter, more constrained utterances (as shown in Table~\ref{tab:dataset_example}), which offer limited phonetic and semantic variation. Moreover, Speechocean prompts students to repeat predefined sentences, unlike the MultiPA data, which includes free-form speech. As a result, both the pause cues between words and the semantic content of the transcripts offer weaker indicators of language proficiency, thereby reducing the effectiveness of TextPA. These dataset differences may also explain the performance inconsistency between Gemini and GPT across the two datasets. Nevertheless, TextPA remains competitive on Speechocean. Note that TextPA relies solely on text tokens, whereas Gemini-2.0-flash-audio uses text tokens for instructions and audio tokens for input speech signals\footnote{The cost of \textit{gemini-2.0-flash} is 0.1 per 1M text tokens and \$0.7 per 1M audio tokens, making Gemini-2.0-flash-audio approximately 3.5 times more expensive in API calls than running TextPA \textit{(Gemini-2.0-flash)} on the Speechocean.}. We also include in-domain models' performance as references. Since TextPA is a zero-shot approach without using training data, the in-domain models naturally perform better. Directly combining the predictions as done with MultiPA data does not lead to improvements for the in-domain setting due to the performance gap. Further investigation is needed to explore more effective ways of leveraging TextPA for in-domain models.

\begin{table}[htbp!]
\centering
\renewcommand{\arraystretch}{1.1}
\begin{tabular}{|c|c|c|}
\hline
\multicolumn{1}{|c|}{}           & \multicolumn{1}{c|}{Accuracy} & Fluency \\ \hline \hline
\multicolumn{3}{|c|}{\cellcolor[HTML]{EFEFEF}Zero-shot}                                           \\ \hline
\multicolumn{1}{|c|}{\begin{tabular}[c]{@{}c@{}}TextPA \\ \textit{(gpt-4o-mini)}\end{tabular}}      & \multicolumn{1}{c|}{0.507} & 0.466 \\ \hline
\multicolumn{1}{|c|}{\begin{tabular}[c]{@{}c@{}}TextPA \\ \textit{(gemini-2.0-flash)}\end{tabular}} & \multicolumn{1}{c|}{0.532} & \textbf{0.557} \\ \hline \hline
\multicolumn{1}{|c|}{Gemini-2.0-flash-audio}                                               & \multicolumn{1}{c|}{\textbf{0.562}} & 0.556 \\ \hline  \hline
\multicolumn{3}{|c|}{\cellcolor[HTML]{EFEFEF}In-domain}                                           \\ \hline
\multicolumn{1}{|c|}{~\cite{lin2023exploiting}} & \multicolumn{1}{c|}{0.72}     & -       \\ \hline
\multicolumn{1}{|c|}{~\cite{liu2023asr}} & \multicolumn{1}{c|}{-}        & 0.795   \\ \hline
\multicolumn{1}{|c|}{MultiPA model}    & \multicolumn{1}{c|}{0.705}    & 0.772   \\ \hline
\end{tabular}
  \caption{Model performance on Speechocean.}
\label{tab:performance_speechocean}
\end{table}

\subsection{Ablation Study on Textual Descriptions of Speech Signals}

\begin{table}[t]
\centering
\renewcommand{\arraystretch}{1.1}
\begin{tabular}{|ccc|}
\hline
\multicolumn{3}{|c|}{\cellcolor[HTML]{EFEFEF}MultiPA data} \\ \hline
\multicolumn{1}{|c|}{} &
  \multicolumn{1}{c|}{Accuracy} &
  Fluency \\ \hline
\multicolumn{1}{|c|}{\begin{tabular}[c]{@{}c@{}}TextPA\\ \textit{(gpt-4o-mini)}\end{tabular}} &
  \multicolumn{1}{c|}{0.728} &
  {\color[HTML]{333333} 0.650} \\ \hline \hline
\multicolumn{1}{|c|}{LLM: all} &
  \multicolumn{1}{c|}{0.643} &
  0.650 \\ \hline
\multicolumn{1}{|c|}{LLM: trans.+cmu} &
  \multicolumn{1}{c|}{0.491} &
  0.485 \\ \hline
\multicolumn{1}{|c|}{LLM: trans.+ipa} &
  \multicolumn{1}{c|}{0.452} &
  0.410 \\ \hline
\multicolumn{1}{|c|}{LLM: transcript} &
  \multicolumn{1}{c|}{0.404} &
  0.432 \\ \hline \hline
\multicolumn{1}{|c|}{IPA match scoring} &
  \multicolumn{1}{c|}{0.653} &
  - \\ \hline
\multicolumn{1}{|c|}{{\color[HTML]{9B9B9B} CMU match scoring}} &
  \multicolumn{1}{c|}{{\color[HTML]{9B9B9B} 0.208}} &
  {\color[HTML]{9B9B9B} -} \\ \hline \hline
\multicolumn{3}{|c|}{\cellcolor[HTML]{EFEFEF}Speechocean} \\ \hline 
\multicolumn{1}{|c|}{} &
  \multicolumn{1}{c|}{Accuracy} &
  Fluency \\ \hline
\multicolumn{1}{|c|}{\begin{tabular}[c]{@{}c@{}}TextPA \\ \textit{(gemini-2.0-flash)}\end{tabular}} &
  \multicolumn{1}{c|}{0.532} &
  0.557 \\ \hline \hline
\multicolumn{1}{|c|}{LLM: all} &
  \multicolumn{1}{c|}{0.456} &
  0.557 \\ \hline
\multicolumn{1}{|c|}{LLM: trans.+cmu} &
  \multicolumn{1}{c|}{0.427} &
  0.553 \\ \hline
\multicolumn{1}{|c|}{LLM: trans.+ipa} &
  \multicolumn{1}{c|}{0.448} &
  0.458 \\ \hline
\multicolumn{1}{|c|}{LLM: transcript} &
  \multicolumn{1}{c|}{0.313} &
  0.310 \\ \hline \hline
\multicolumn{1}{|c|}{IPA match scoring} &
  \multicolumn{1}{c|}{0.507} &
  - \\ \hline
\multicolumn{1}{|c|}{{\color[HTML]{9B9B9B} CMU match scoring}} &
  \multicolumn{1}{c|}{{\color[HTML]{9B9B9B} 0.263}} &
  {\color[HTML]{9B9B9B} -} \\ \hline
\end{tabular}
  \caption{Ablation study of text-based acoustic cues. We selected the LLM with the best performance on each dataset as the representative model: \textit{gpt-4o-mini} for the MultiPA data and \textit{gemini-2.0-flash} for the Speechocean data. \textit{LLM: transcript} uses only the transcript as input. \textit{LLM: trans.+ ipa} and \textit{trans.+ cmu} add IPA or CMU sequences, respectively. \textit{LLM: all} combines all three inputs: transcript, IPA, and CMU. Note that the fluency scores for \textit{LLM: all} and TextPA are identical, as IPA score matching is only used to refine accuracy.}
\label{tab:ablation_acoustic}
\end{table}

First, we evaluated the performance of accuracy scoring based on phoneme sequence matching (Table~\ref{tab:ablation_acoustic}). Our findings demonstrate that IPA match scoring is a straightforward yet highly effective method for assessing pronunciation accuracy. We also investigated the performance of CMU match scoring. Similar to IPA match scoring, the words in the transcript are mapped to CMU labels using the dictionary, and then compared with the recognized CMU sequence through normalized Smith-Waterman similarity scores. However, the results indicate that the CMU sequence is less effective for accuracy assessment compared to the IPA sequence. This difference may stem from the greater phonetic detail provided by the IPA, which contains more than 107 syllable letters, while the CMU set contains only 39 phonemes. 

Table~\ref{tab:ablation_acoustic} also reports an ablation study evaluating which textual descriptions of acoustic cues are most effective for language models in pronunciation assessment. When using an LLM, the transcript alone can offer insights. Augmenting the input with recognized IPA sequences improves performance, particularly in accuracy, as the LLM can compare word transcriptions with their phonetic transcriptions to better identify mispronunciations. Adding CMU sequences alongside the transcript helps to enhance both accuracy and fluency as well: accuracy improves for similar reasons as with IPA, while fluency benefits from the pause information encoded in CMU sequences. Overall, combining the transcript, CMU, and IPA sequences leads to the best performance, with IPA match scoring providing additional boosts in accuracy.

\subsection{Impact of ASR Transcription Quality}

Transcripts play a crucial role in TextPA. To examine the affect of ASR model quality (i.e., transcription quality), we compared LLM-based assessment using transcripts generated by two Whisper variants: \textit{large-v3-en} (denoted as \textit{large-en}) and \textit{tiny}. The \textit{large-en} model, with 1550M parameters, is English-only and generates higher-quality transcripts that are more robust to inaccurate pronunciation. In contrast, the \textit{tiny} model, with only 39M parameters and multilingual training, is more likely to produce transcription errors or misclassify English as a different language when pronunciation is inaccurate. 

As shown in Table~\ref{tab:transcript_analysis}, when transcripts alone are used as input to the LLM, \textit{tiny} yields better assessment results than \textit{large-en}. This observation can be illustrated through an analogy: using \textit{large-en} is like speaking to a listener with excellent English comprehension -- they can understand you even if your pronunciation is poor. In contrast, the \textit{tiny} model resembles a listener with limited English ability, who can only understand clearly articulated speech. Whether a person with strong English listening comprehension (i.e., \textit{large-en}) can understand you provides less insight into your pronunciation. In contrast, if people with weaker listening ability (i.e., \textit{tiny}) can understand you easily, it indicates that your pronunciation is good.

Although the transcripts from \textit{tiny} models perform better on their own, the \textit{large-en} model is more effective within the TextPA framework. In TextPA, we incorporate the IPA and CMU sequences along with the transcript. Inaccurate pronunciation can lead to unnatural IPA and CMU sequences, offering similar insights to the transcript of \textit{tiny} model. In addition, because the transcript serves as a baseline for comparison, excessive ASR errors introduce noise that reduces reliability. Overall, we believe that a stronger ASR model, such as \textit{large-en}, is the better choice within the TextPA structure.

\begin{table}[htbp!]
\centering
\renewcommand{\arraystretch}{1.1}
\resizebox{0.97\linewidth}{!}{%
\begin{tabular}{|ccccc|}
\hline
\multicolumn{1}{|c|}{} &
  \multicolumn{2}{c|}{\cellcolor[HTML]{EFEFEF}Accuracy} &
  \multicolumn{2}{c|}{Fluency} \\ \hline
\multicolumn{1}{|c|}{} &
  \multicolumn{1}{c|}{\cellcolor[HTML]{EFEFEF}\textit{large-en}} &
  \multicolumn{1}{c|}{\cellcolor[HTML]{EFEFEF}\textit{tiny}} &
  \multicolumn{1}{c|}{\textit{large-en}} &
  \textit{tiny} \\ \hline
\multicolumn{5}{|c|}{MultiPA data} \\ \hline
\multicolumn{1}{|c|}{\begin{tabular}[c]{@{}c@{}}LLM: all\\ \textit{(gpt-4o-mini)}\end{tabular}} &
  \multicolumn{1}{c|}{\cellcolor[HTML]{EFEFEF}\textbf{0.643}} &
  \multicolumn{1}{c|}{\cellcolor[HTML]{EFEFEF}0.569} &
  \multicolumn{1}{c|}{\textbf{0.650}} &
  0.546 \\ \hline
\multicolumn{1}{|c|}{LLM: transcript} &
  \multicolumn{1}{c|}{\cellcolor[HTML]{EFEFEF}0.404} &
  \multicolumn{1}{c|}{\cellcolor[HTML]{EFEFEF}\textbf{0.556}} &
  \multicolumn{1}{c|}{0.432} &
  \textbf{0.442} \\ \hline
\multicolumn{5}{|c|}{Speechocean} \\ \hline
\multicolumn{1}{|c|}{\begin{tabular}[c]{@{}c@{}}LLM: all\\ \textit{(gemini-2.0-flash)}\end{tabular}} &
  \multicolumn{1}{c|}{\cellcolor[HTML]{EFEFEF}0.456} &
  \multicolumn{1}{c|}{\cellcolor[HTML]{EFEFEF}\textbf{0.481}} &
  \multicolumn{1}{c|}{\textbf{0.557}} &
  0.523 \\ \hline
\multicolumn{1}{|c|}{LLM: transcript} &
  \multicolumn{1}{c|}{\cellcolor[HTML]{EFEFEF}0.313} &
  \multicolumn{1}{c|}{\cellcolor[HTML]{EFEFEF}\textbf{0.409}} &
  \multicolumn{1}{c|}{0.310} &
  \textbf{0.431} \\ \hline
\end{tabular}%
}
  \caption{Impact of ASR transcription quality.}
\label{tab:transcript_analysis}
\end{table}

\subsection{Analysis of Basic vs.~Detailed Scoring Guidelines}

We investigated the impact of providing different instructions to the LLM, including basic and detailed scoring guidelines (Table~\ref{tab:instruction_analysis}). The basic scoring guideline prompts the LLM to assign \textit{a scoring range (1-5), where a higher score indicates better pronunciation, with a score of 5 reflecting native-speaker proficiency.} The detailed scoring guideline, on the other hand, provides the same detailed annotation guidelines used by human annotators. The detailed guidelines define the language proficiency for each score level. For example, for MultiPA data, an accuracy score of 5 means \textit{“Excellent: The overall pronunciation is nearly perfect with accurate articulation of all sounds,”} while a score of 4 means \textit{“Good: Minor pronunciation errors may be present, but overall, the pronunciation is highly accurate and easily understandable”}, and so on. Results suggest that the effectiveness is dataset-dependent, possibly influenced by how the guidelines are written. However, incorporating a detailed scoring guideline has the potential to reduce performance, while also lengthening the input text prompt and increasing model operating costs.

\begin{table}[htbp!]
\centering
\renewcommand{\arraystretch}{1.1}
\resizebox{0.97\linewidth}{!}{%
\begin{tabular}{|ccccc|}
\hline
\multicolumn{1}{|c|}{} &
  \multicolumn{2}{c|}{\cellcolor[HTML]{EFEFEF}Accuracy} &
  \multicolumn{2}{c|}{Fluency} \\ \hline
\multicolumn{1}{|c|}{} &
  \multicolumn{1}{c|}{\cellcolor[HTML]{EFEFEF}Basic} &
  \multicolumn{1}{c|}{\cellcolor[HTML]{EFEFEF}Detailed} &
  \multicolumn{1}{c|}{Basic} &
  Detailed \\ \hline
\multicolumn{5}{|c|}{MultiPA data} \\ \hline
\multicolumn{1}{|c|}{\begin{tabular}[c]{@{}c@{}}LLM: all\\ \textit{(gpt-4o-mini)}\end{tabular}} &
  \multicolumn{1}{c|}{\cellcolor[HTML]{EFEFEF}\textbf{0.643}} &
  \multicolumn{1}{c|}{\cellcolor[HTML]{EFEFEF}0.500} &
  \multicolumn{1}{c|}{\textbf{0.650}} &
  0.543 \\ \hline
\multicolumn{1}{|c|}{\begin{tabular}[c]{@{}c@{}}LLM: all\\ \textit{(gemini-2.0-flash)}\end{tabular}} &
  \multicolumn{1}{c|}{\cellcolor[HTML]{EFEFEF}0.554} &
  \multicolumn{1}{c|}{\cellcolor[HTML]{EFEFEF}\textbf{0.596}} &
  \multicolumn{1}{c|}{\textbf{0.556}} &
  0.499 \\ \hline
\multicolumn{5}{|c|}{Speechocean} \\ \hline
\multicolumn{1}{|c|}{\begin{tabular}[c]{@{}c@{}}LLM: all\\ \textit{(gpt-4o-mini)}\end{tabular}} &
  \multicolumn{1}{c|}{\cellcolor[HTML]{EFEFEF}0.420} &
  \multicolumn{1}{c|}{\cellcolor[HTML]{EFEFEF}\textbf{0.474}} &
  \multicolumn{1}{c|}{0.466} &
  \textbf{0.544} \\ \hline
\multicolumn{1}{|c|}{\begin{tabular}[c]{@{}c@{}}LLM: all\\ \textit{(gemini-2.0-flash)}\end{tabular}} &
  \multicolumn{1}{c|}{\cellcolor[HTML]{EFEFEF}0.456} &
  \multicolumn{1}{c|}{\cellcolor[HTML]{EFEFEF}\textbf{0.470}} &
  \multicolumn{1}{c|}{0.557} &
  \textbf{0.561} \\ \hline
\end{tabular}%
}
  \caption{Performance with basic or detailed guidelines.}
\label{tab:instruction_analysis}
\end{table}

\section{Analysis of TextPA Reasoning}

Table~\ref{tab:reasoning_example} presents examples of TextPA reasoning. We observed that \emph{gemini-2.0-flash} is more likely to refer to the detailed CMU and IPA phoneme sequences to support its assessment, while \emph{gpt-4o-mini} tends to provide more generalized observations. Also, \emph{gpt-4o-mini} often frames its evaluation as verifying whether the phoneme sequence aligns with the pronunciation, focusing on the “phoneme sequence” or the “ASR system”  rather than commenting directly on pronunciation.

\begin{table}[htbp!]
\centering
  \includegraphics[scale=0.8]{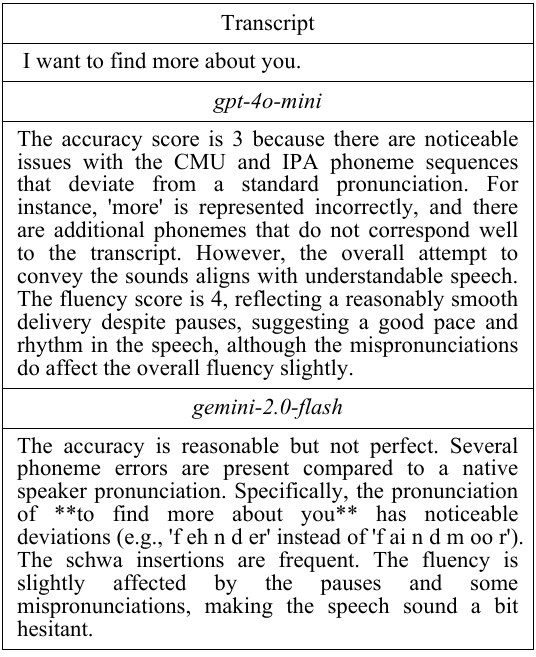} 
  \caption {Examples of TextPA reasoning.}
    \label{tab:reasoning_example}
\end{table}

To gain deeper insights, we manually annotated the TextPA reasoning into four categories: hallucination, correct, constructive, and irrelevant. Hallucination refers to cases where the reasoning clearly misrepresents the audio, such as stating \emph{“closely matches standard native speaker articulation”} when the pronunciation is clearly non-standard. Correct indicates reasoning that aligns with the audio but does not provide actionable details; for example, \emph{“The accuracy score of 3 reflects a moderate level of pronunciation correctness. While there are identifiable phonetic errors, the core message is still comprehensible.”} Constructive reasoning identifies specific areas for improvement, such as \emph{“**calls us** is recognized as **cars**”} Finally, irrelevant refers to reasoning that is unrelated to pronunciation, such as comments on grammar or the transcript.

\begin{figure}[htbp!]
\centering
  \includegraphics[scale=0.85]{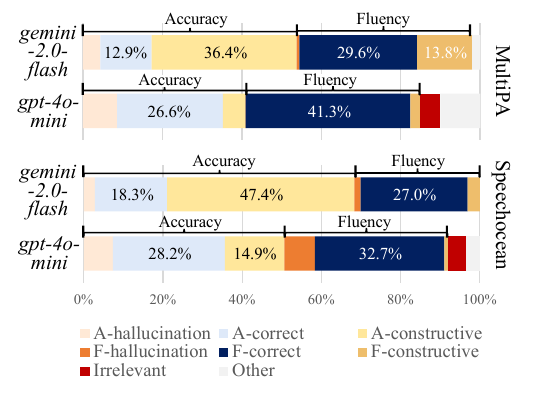} 
  \caption {Coverage analysis of TextPA reasoning. \emph{“A”} denotes accuracy, and \emph{“F”} denotes fluency.}
    \label{fig:description_analysis}
\end{figure}

We then measured the coverage of each category in the TextPA-generated reasoning (Figure~\ref{fig:description_analysis}). Coverage was determined by tokenizing the reasoning descriptions and calculating the proportion of tokens belonging to each category. For MultiPA data, 53\% of \emph{gemini-2.0-flash}’s generated descriptions relate to accuracy and 44\% to fluency, while 40\% of \emph{gpt-4o-mini}’s descriptions relate to accuracy and 43\% to fluency. \emph{Gemini-2.0-flash} allocates a greater proportion of content to accuracy than to fluency, whereas \emph{gpt-4o-mini}’s content is more evenly split. The rest contains irrelevant reasoning or general overviews of pronunciation proficiency. For Speechocean, we randomly selected 25 samples for annotation. Compared to MultiPA data, both LLMs place considerably greater emphasis on accuracy than on fluency on Speechocean, with 68\% vs. 31\% for Gemini, and 50\% vs. 41\% for GPT. This difference is likely due to the shorter utterances in Speechocean, which provide limited material to observe natural speech flow or identify disruptions, making fluency assessment less feasible. On both datasets, \emph{gemini-2.0-flash} generally provides more constructive reasoning compared to \emph{gpt-4o-mini}. For both LLMs, constructive reasoning occurs more frequently for accuracy than for fluency, likely because accuracy is more clearly defined and can be evaluated more objectively. Overall, roughly 76\% of the context in \emph{gpt-4o-mini}-based TextPA reasoning is either correct or constructive, while over 90\% of \emph{gemini-2.0-flash}-based reasoning falls into these categories, highlighting the strong potential of TextPA to generate meaningful descriptive feedback for pronunciation assessment.

\section{Background}
\subsection{Speech Pronunciation Assessment}

Speech pronunciation assessment models can be categorized into closed- or open-response scenarios. In closed-response settings, learners read a predetermined sentence, which serves as the ground-truth transcript for the model to guide the assessment. A common approach in this scenario extracted Goodness of Pronunciation (GoP) features to train an acoustic model~\cite{gong2022transformer, do2023hierarchical}. In addition to GoP, various other features have been explored for model training, including acoustic embeddings from self-supervised learning (SSL) models, prosodic features such as duration and energy, and transcript-based features such as word embeddings~\cite{chao20223m, yan2025conpco}. In ~\cite{wu2025integrating}, researchers fine-tuned an LLM using audio tokens and text prompts to provide feedback on phone errors. However, the performance of models trained with ground-truth transcripts may degrade significantly when such transcripts are unavailable. On the other hand, open-response scenarios allow learners to speak freely or respond to prompts, enabling a more authentic evaluation of their pronunciation skills. Models designed for open-response tasks do not rely on ground-truth transcripts. Instead, they leverage ASR outputs or avoid ASR entirely~\cite{lin2021deep, kim2022automatic, chen2023multipa, liu2023asr}. Most prior studies rely on audio-score pair data to train acoustic models for pronunciation assessment, whereas zero-shot approaches have been largely unexplored. In~\cite{liu2023zero}, researchers scored pronunciation based on the number of incorrectly recovered tokens from an SSL model. However, like other previous studies, it provided only numerical feedback instead of more interpretable or explainable assessments.

\subsection{LLM for Language Learning}

LLMs have had a significant impact on education, with many studies exploring how tools like ChatGPT can support language learning~\cite{lo2024exploring, c2023impact}. These models have proven effective in helping learners identify and correct writing errors, improve the quality of their writing~\cite{barrot2023using}, and receive automated feedback~\cite{mizumoto2023exploring}. Few studies have focused on using LLMs to support speaking skills. \cite{kim2023young} used ChatGPT as a conversational partner in role-playing tasks, while \cite{lee2023visionary} used it to generate topics for oral practice. A study by~\cite{wang2023assessing} used ChatGPT to assess how well ESL learners placed pauses in their speech. However, the potential of LLMs to support other aspects of oral language skills, such as pronunciation accuracy and fluency as in TextPA, remains underexplored.

\section{Conclusion}

We propose TextPA, a zero-shot pronunciation assessment method that leverages interpretable, textual representations of speech signals to assess pronunciation accuracy and fluency. These descriptions include transcripts, IPA, and CMU phoneme sequences, collectively reflecting pronunciation characteristics. Specifically, semantically unnatural transcripts may signal pronunciation issues, mismatches between canonical and recognized phoneme sequences reflect articulation errors, and inappropriate pauses embedded in CMU sequences reveal disfluencies. Experimental results demonstrate that LLMs can effectively leverage textual description of speech to assess different aspects of pronunciation. Unlike conventional models trained on audio-score pairs, TextPA operates without supervision. TextPA focuses on human-readable representations and prior knowledge of pronunciation, aiming to provide interpretable and explainable feedback that go beyond a score. We hope this work offers a new perspective on pronunciation assessment. Building on our initial exploration, future research could further develop methods to more effectively integrate TextPA with audio-trained models, combining their strengths to improve assessment accuracy and feedback quality for learners.

\section*{Limitations}

While prosody is an important aspect of pronunciation, we found it difficult to effectively assess using our text-based approach. Compared to accuracy and fluency, prosodic features such as rhythm and intonation are harder to describe precisely in written form, making them less suitable for methods that rely solely on textual representations. As a result, the LLM struggled to reliably evaluate prosody without compromising assessment performance on accuracy and fluency. In addition, both the LLM and the ASR system introduce variability across runs, leading to inconsistent assessment results. In addition, budget constraints limited our ability to use the most advanced LLMs or to evaluate large ALMs across all settings. Lastly, the LLM occasionally produces hallucinations or content irrelevant to the reasoning. While most outputs align with the audio, providing more practically actionable feedback could better support learners. Exhaustive manual review of reasoning results is beyond the scope of this study, and no established metric currently exists to automatically verify correctness. Further investigation is needed to determine the conditions under which the LLM is more likely to generate errors and to develop strategies that both prevent such errors and enhance actionable feedback. These limitations suggest future work in prosody modeling, dataset expansion, and automatic reasoning evaluation.

Although certain words may have multiple valid pronunciations depending on the speaker’s accent, our study did not consider accent variation, since the majority of the data involved attempts to mimic General American English. Consequently, a potential risk of this study is an overemphasis on a single accent. While many English learners aim to emulate native speakers, the more practical goal in everyday communication is to express one’s opinions clearly and be understood. This highlights the importance of balancing pronunciation assessment systems between intelligibility and nativeness. When such systems overemphasize native-like pronunciation, which is often tied to a specific accent, they might erroneously mark understandable speech as “wrong.” Failing to strike this balance can marginalize learners’ linguistic identities and encourage unnecessary \textit{accent reduction} at the expense of communicative effectiveness. In addition, an overly narrow model can reinforce the idea that only a single variety of English is valid, thereby undermining the rich diversity of global English accents.

\bibliography{custom}

\begin{thebibliography}{31}
\providecommand{\natexlab}[1]{#1}

\bibitem[{Barrot(2023)}]{barrot2023using}
Jessie~S Barrot. 2023.
\newblock Using {ChatGPT} for second language writing: Pitfalls and potentials.
\newblock \emph{Assessing Writing}, 57:100745.

\bibitem[{Beckman and Hirschberg(1994)}]{beckman1994tobi}
Mary~E Beckman and Julia Hirschberg. 1994.
\newblock The tobi annotation conventions.
\newblock \emph{Ohio State University}.

\bibitem[{Bernard and Titeux(2021)}]{Bernard2021}
Mathieu Bernard and Hadrien Titeux. 2021.
\newblock \href {https://doi.org/10.21105/joss.03958} {Phonemizer: Text to phones transcription for multiple languages in python}.
\newblock \emph{Journal of Open Source Software}, 6(68):3958.

\bibitem[{C~Meniado(2023)}]{c2023impact}
Joel C~Meniado. 2023.
\newblock The impact of {ChatGPT} on english language teaching, learning, and assessment: A rapid review of literature.
\newblock \emph{Arab World English Journals (AWEJ) Volume}, 14.

\bibitem[{Chao et~al.(2022)Chao, Lo, Wu, Sung, and Chen}]{chao20223m}
Fu-An Chao, Tien-Hong Lo, Tzu-I Wu, Yao-Ting Sung, and Berlin Chen. 2022.
\newblock {3M}: An effective multi-view, multi-granularity, and multi-aspect modeling approach to english pronunciation assessment.
\newblock In \emph{Proc. APSIPA ASC 2022}, pages 575--582. IEEE.

\bibitem[{Chen et~al.(2024)Chen, Yu, and Hirschberg}]{chen2023multipa}
Yu-Wen Chen, Zhou Yu, and Julia Hirschberg. 2024.
\newblock {MultiPA}: a multi-task speech pronunciation assessment model for open response scenarios.
\newblock In \emph{Proc. INTERSPEECH 2024}, pages 297--301.

\bibitem[{Chu et~al.(2023)Chu, Xu, Zhou, Yang, Zhang, Yan, Zhou, and Zhou}]{chu2023qwen}
Yunfei Chu, Jin Xu, Xiaohuan Zhou, Qian Yang, Shiliang Zhang, Zhijie Yan, Chang Zhou, and Jingren Zhou. 2023.
\newblock Qwen-audio: Advancing universal audio understanding via unified large-scale audio-language models.
\newblock \emph{arXiv preprint arXiv:2311.07919}.

\bibitem[{Deshmukh et~al.(2024)Deshmukh, Alharthi, Elizalde, Gamper, Ismail, Singh, Raj, and Wang}]{deshmukh2024pam}
Soham Deshmukh, Dareen Alharthi, Benjamin Elizalde, Hannes Gamper, Mahmoud~Al Ismail, Rita Singh, Bhiksha Raj, and Huaming Wang. 2024.
\newblock {PAM}: Prompting audio-language models for audio quality assessment.
\newblock In \emph{Proc. INTERSPEECH 2024}.

\bibitem[{Do et~al.(2023)Do, Kim, and Lee}]{do2023hierarchical}
Heejin Do, Yunsu Kim, and Gary~Geunbae Lee. 2023.
\newblock Hierarchical pronunciation assessment with multi-aspect attention.
\newblock In \emph{Proc. ICASSP 2023}. IEEE.

\bibitem[{Elizalde et~al.(2023)Elizalde, Deshmukh, Al~Ismail, and Wang}]{elizalde2023clap}
Benjamin Elizalde, Soham Deshmukh, Mahmoud Al~Ismail, and Huaming Wang. 2023.
\newblock {CLAP}: Learning audio concepts from natural language supervision.
\newblock In \emph{Proc. ICASSP 2023}. IEEE.

\bibitem[{Gong et~al.(2022)Gong, Chen, Chu, Chang, and Glass}]{gong2022transformer}
Yuan Gong, Ziyi Chen, Iek-Heng Chu, Peng Chang, and James Glass. 2022.
\newblock Transformer-based mmulti-aspect multi-granularity non-native english speaker ppronunciation assessment.
\newblock In \emph{Proc. ICASSP 2022}, pages 7262--7266. IEEE.

\bibitem[{Kim et~al.(2022)Kim, Jeon, Seo, and Kim}]{kim2022automatic}
Eesung Kim, Jae-Jin Jeon, Hyeji Seo, and Hoon Kim. 2022.
\newblock Automatic pronunciation assessment using self-supervised speech representation learning.
\newblock In \emph{Proc. INTERSPEECH 2022}, pages 1411--1415.

\bibitem[{Kim and Park(2023)}]{kim2023young}
Sol Kim and Seon-Ho Park. 2023.
\newblock Young korean {EFL} learners' perception of role-playing scripts: {ChatGPT} vs. textbooks.
\newblock \emph{Journal of English Language and Linguistics}, 23:1136--1153.

\bibitem[{Lee et~al.(2023)Lee, Hsia, Tsoy, Choi, Hou, and Ni}]{lee2023visionary}
Hyungmin Lee, Chen-Chun Hsia, Aleksandr Tsoy, Sungmin Choi, Hanchao Hou, and Shiguang Ni. 2023.
\newblock {VisionARy}: Exploratory research on contextual language learning using {AR} glasses with {ChatGPT}.
\newblock In \emph{Proceedings of the 15th biannual conference of the Italian SIGCHI chapter}, pages 1--6.

\bibitem[{Lin and Wang(2021)}]{lin2021deep}
Binghuai Lin and Liyuan Wang. 2021.
\newblock Deep feature transfer learning for automatic pronunciation assessment.
\newblock In \emph{Proc. INTERSPEECH 2021}, pages 4438--4442.

\bibitem[{Lin and Wang(2022)}]{lin2023exploiting}
Binghuai Lin and Liyuan Wang. 2022.
\newblock Exploiting information from native data for non-native automatic pronunciation assessment.
\newblock In \emph{Proc. SLT 2022}, pages 708--714. IEEE.

\bibitem[{Liu et~al.(2023{\natexlab{a}})Liu, Shi, and Wang}]{liu2023zero}
Hongfu Liu, Mingqian Shi, and Ye~Wang. 2023{\natexlab{a}}.
\newblock Zero-shot automatic pronunciation assessment.
\newblock In \emph{Proc. INTERSPEECH 2023}, pages 1009--1013.

\bibitem[{Liu et~al.(2023{\natexlab{b}})Liu, Fu, Tian, Shi, Li, Ma, and Lee}]{liu2023asr}
Wei Liu, Kaiqi Fu, Xiaohai Tian, Shuju Shi, Wei Li, Zejun Ma, and Tan Lee. 2023{\natexlab{b}}.
\newblock An {ASR-free} fluency scoring approach with self-supervised learning.
\newblock In \emph{Proc. ICASSP 2023}. IEEE.

\bibitem[{Lo et~al.(2024)Lo, Yu, Xu, Ng, and Jong}]{lo2024exploring}
Chung~Kwan Lo, Philip Leung~Ho Yu, Simin Xu, Davy Tsz~Kit Ng, and Morris Siu-yung Jong. 2024.
\newblock Exploring the application of {ChatGPT} in {ESL/EFL} education and related research issues: a systematic review of empirical studies.
\newblock \emph{Smart Learning Environments}, 11(1):50.

\bibitem[{Mizumoto and Eguchi(2023)}]{mizumoto2023exploring}
Atsushi Mizumoto and Masaki Eguchi. 2023.
\newblock Exploring the potential of using an {AI} language model for automated essay scoring.
\newblock \emph{Research Methods in Applied Linguistics}, 2(2):100050.

\bibitem[{Radford et~al.(2023)Radford, Kim, Xu, Brockman, McLeavey, and Sutskever}]{radford2023robust}
Alec Radford, Jong~Wook Kim, Tao Xu, Greg Brockman, Christine McLeavey, and Ilya Sutskever. 2023.
\newblock Robust speech recognition via large-scale weak supervision.
\newblock In \emph{Proc. ICML 2023}, pages 28492--28518.

\bibitem[{Smith et~al.(1981)Smith, Waterman et~al.}]{smith1981identification}
Temple~F Smith, Michael~S Waterman, and 1 others. 1981.
\newblock Identification of common molecular subsequences.
\newblock \emph{Journal of molecular biology}, 147(1):195--197.

\bibitem[{Tang et~al.(20234)Tang, Yu, Sun, Chen, Tan, Li, Lu, Ma, and Zhang}]{tang2023salmonn}
Changli Tang, Wenyi Yu, Guangzhi Sun, Xianzhao Chen, Tian Tan, Wei Li, Lu~Lu, Zejun Ma, and Chao Zhang. 20234.
\newblock {SALMONN}: Towards generic hearing abilities for large language models.
\newblock \emph{Proc. ICLR 2024}.

\bibitem[{Wang et~al.(2025{\natexlab{a}})Wang, He, Liu, Deng, Wei, and Zhao}]{wang2025exploring}
Ke~Wang, Lei He, Kun Liu, Yan Deng, Wenning Wei, and Sheng Zhao. 2025{\natexlab{a}}.
\newblock Exploring the potential of large multimodal models as effective alternatives for pronunciation assessment.
\newblock \emph{arXiv preprint arXiv:2503.11229}.

\bibitem[{Wang et~al.(2025{\natexlab{b}})Wang, Yu, Yang, Tang, Li, Zhuang, Chen, Tian, Zhang, Sun et~al.}]{wang2025enabling}
Siyin Wang, Wenyi Yu, Yudong Yang, Changli Tang, Yixuan Li, Jimin Zhuang, Xianzhao Chen, Xiaohai Tian, Jun Zhang, Guangzhi Sun, and 1 others. 2025{\natexlab{b}}.
\newblock Enabling auditory large language models for automatic speech quality evaluation.
\newblock In \emph{Proc. ICASSP 2025}. IEEE.

\bibitem[{Wang et~al.(2023)Wang, Mao, Wu, Xia, Deng, and Tien}]{wang2023assessing}
Zhiyi Wang, Shaoguang Mao, Wenshan Wu, Yan Xia, Yan Deng, and Jonathan Tien. 2023.
\newblock Assessing phrase break of {ESL} speech with pre-trained language models and large language models.
\newblock In \emph{Proc. INTERSPEECH 2023}, pages 4194--4198.

\bibitem[{Wu et~al.(2025)Wu, Xu, Chen, and Meng}]{wu2025integrating}
Minglin Wu, Jing Xu, Xueyuan Chen, and Helen Meng. 2025.
\newblock Integrating potential pronunciations for enhanced mispronunciation detection and diagnosis ability in llms.
\newblock In \emph{Proc. ICASSP 2025}. IEEE.

\bibitem[{Xu et~al.(2021)Xu, Baevski, and Auli}]{xu2021simple}
Qiantong Xu, Alexei Baevski, and Michael Auli. 2021.
\newblock Simple and effective zero-shot cross-lingual phoneme recognition.
\newblock In \emph{Proc. INTERSPEECH 2021}, pages 2113--2117.

\bibitem[{Yan et~al.(2025)Yan, Wang, Li, Lin, Wang, Chao, and Chen}]{yan2025conpco}
Bi-Cheng Yan, Yi-Cheng Wang, Jiun-Ting Li, Meng-Shin Lin, Hsin-Wei Wang, Wei-Cheng Chao, and Berlin Chen. 2025.
\newblock {ConPCO}: Preserving phoneme characteristics for automatic pronunciation assessment leveraging contrastive ordinal regularization.
\newblock In \emph{Proc. ICASSP 2025}. IEEE.

\bibitem[{Zhang et~al.(2021)Zhang, Zhang, Wang, Yan, Song, Huang, Li, Povey, and Wang}]{zhang2021speechocean762}
Junbo Zhang, Zhiwen Zhang, Yongqing Wang, Zhiyong Yan, Qiong Song, Yukai Huang, Ke~Li, Daniel Povey, and Yujun Wang. 2021.
\newblock Speechocean762: An open-source non-native {English} speech corpus for pronunciation assessment.
\newblock In \emph{Proc. INTERSPEECH 2021}, pages 3710--3714.

\bibitem[{Zhu et~al.(2022)Zhu, Zhang, and Jurgens}]{zhu2022phone}
Jian Zhu, Cong Zhang, and David Jurgens. 2022.
\newblock Phone-to-audio alignment without text: A semi-supervised approach.
\newblock In \emph{Proc. ICASSP 2022}, pages 8167--8171. IEEE.

\end{thebibliography}

\appendix

\section{Prompt}

\begin{figure}[t]
\centering
\includegraphics[scale=0.85]{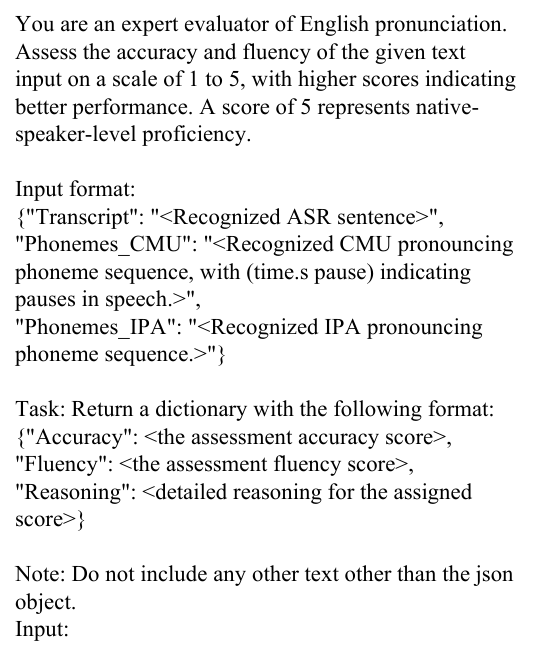}
  \caption{LLM prompt.}
  \label{fig:prompt_textPA}
\end{figure}

Figure~\ref{fig:prompt_textPA} shows the TextPA prompt for LLM; ALM prompt follows a similar format, but does not include input format instructions. We observed that Gemini is more likely to return results that do not match the required format, whereas GPT tends to produce outputs that can be directly saved as JSON files. If the model fails to generate a correctly formatted output for a given test sample, we re-run it until a valid result is obtained.

\section{Prosody assessment}
\label{sec:appendix}

We investigate whether LLM could assess prosody from textual descriptions. We only used the MultiPA data for this part of the study, as most sentences in Speechocean are short and do not contain sufficient prosodic variation for a reliable assessment. First, we prompted the LLM to evaluate prosody in addition to accuracy and fluency. As shown in Table~\ref{tab:performance_prosody_1}, the model performs worse in terms of prosody assessment compared to fluency and accuracy. In addition, introducing prosody as an additional assessment criterion leads to a decrease in the model's performance in both accuracy and fluency.

\begin{table}[htbp!]
\centering
\renewcommand{\arraystretch}{1.1}
\resizebox{0.97\linewidth}{!}{%
\begin{tabular}{|c|c|c|c|}
\hline
 & Accuracy & Fluency & Prosody \\ \hline
\begin{tabular}[c]{@{}c@{}}LLM: all\\ \textit{(gpt-4o-mini)}\end{tabular}                        & 0.633    & 0.678   & -       \\ \hline
\begin{tabular}[c]{@{}c@{}} LLM\(_{p}\): all \\ \textit{(gpt-4o-mini)}\end{tabular} & 0.590    & 0.549   & 0.243   \\ \hline
\end{tabular}%
}
  \caption{LLM performance with and without prosody assessment. \emph{LLM\(_{p}\): all} is \emph{LLM: all} with the introduction of prosody as an additional assessment criterion. Note that the transcript is generated using \textit{turbo} version of Whisper, an optimized version of \textit{large-v3} that provides faster transcription with minimal loss in accuracy. The results indicate that \textit{turbo} performs comparably to \textit{large-v3-en}. (Section~\ref{sec:performance_on_free_speech})}
  \label{tab:performance_prosody_1}
\end{table}

We explore textual descriptions of prosody using annotations from the ToBI (Tones and Break Indices) system~\cite{beckman1994tobi}\footnote{https://github.com/monikaUPF/PyToBI}, which provides a standardized approach to annotate intonation and phrasing patterns in spoken English. ToBI includes two primary components: the break index and the tone index, both of which are crucial for understanding the prosody of speech signals. The break index ranges from 0 to 4 and is defined as follows:


\begin{tabularx}{0.9\linewidth}{lX}
0: & Clear phonetic marks for clitic groups \\
1: & Most phrase-medial word boundaries \\
2: & Strong disjuncture, pause or virtual pause, no tonal marks \\
3: & Intermediate intonation phrase boundary \\
4: & Full intonation phrase boundary \\
\end{tabularx}

The tone index includes the following categories:

\begin{tabularx}{0.9\linewidth}{lX}
H: & High pitch in the local pitch range \\
L: & Low pitch in the local pitch range \\
\textasteriskcentered: & Pitch accent, indicating that the word is stressed \\
\%: & The end of an intonation phrase \\
- or \verb|--|\verb|--|: & A phrase's accent \\
\end{tabularx}

Table~\ref{tab:prosody_results} presents a selection of examples from our attempts to assess prosody using an LLM. The experimental results indicate that the LLM is less effective in assessing prosody, and requiring it to do so leads to a decline performance in accuracy and fluency. A possible reason for this is that prosody is harder to capture accurately using textual descriptions. Since prosody is less commonly expressed in written form, the LLM has more difficulty leveraging its inherent knowledge for prosody assessment.

\begin{table*}[]
  \includegraphics[scale=0.85]{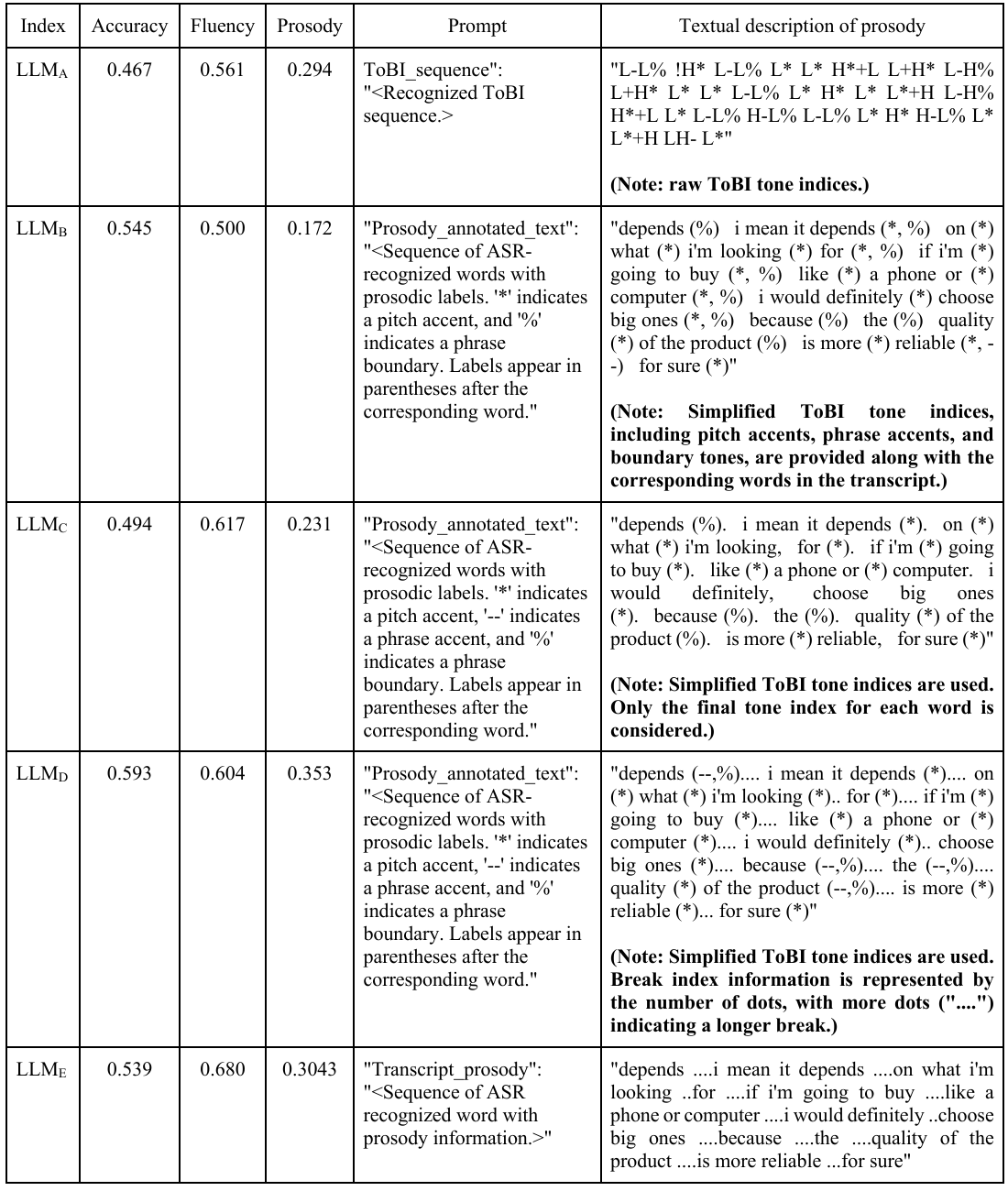} \hfill
  \caption {LLM performance in the presence of textual prosody descriptions. The Prompt column displays the additional instructions given to the LLM, beyond the standard prompt shown in Figure~\ref{fig:prompt_textPA}. The Textual Description of Prosody column illustrates an example input provided to the LLM.}
    \label{tab:prosody_results}
\end{table*}

\end{document}